\documentclass{cjaa}                   

\usepackage{graphicx}                   
\input{epsf.sty}                        
\input{psfig.sty}                       

\begin{document}

   \title{Detecting sub-mJy sources with the EVN
}

   \volnopage{Vol.0 (200x) No.0, 000--000}      
   \setcounter{page}{1}          

   \author{Zsolt Paragi
      \inst{1}
   \and Michael A. Garrett
      \inst{1}
   \and Andrew D. Biggs
      \inst{1}
     } 
   \offprints{Z. Paragi}                   

   \institute{Joint Institute for VLBI in Europe, Postbus 2, 7990 AA Dwingeloo,
             The Netherlands\\
             \email{zparagi@jive.nl}
    } 

   \date{Received~~2004 month day; accepted~~2004~~month day}

   \abstract{ Some microquasars are permanently bright radio sources while others are faint
but produce powerful radio outbursts. Most of the X-ray binaries (XRB) however are very faint 
or undetected 
in the radio regime.  The European VLBI Network (EVN) recently introduced the Mark5 recording
system which allows data rates of up to 1~Gbit/s. This increases the sensitivity of the array
significantly. We briefly describe recent developments in the EVN in terms of reliability of
the network and also data quality. We demonstrate the power of the EVN in detecting sub-mJy
radio sources with modest integration times (order of hours). This high sensitivity capability
will permit the study of variable Galactic sources at milliarcsecond
resolution. An estimate is given for the lowest detectable mass of hypothetical intermediate-mass
black holes (IMBH) in nearby galaxies, provided these are located in radio-jet systems analogous
to microquasars and active galactic nuclei. 
}
   \authorrunning{Z. Paragi, M.A. Garrett \& A.D. Biggs }            
   \titlerunning{Detecting sub-mJy sources with the EVN}  

   \maketitle

%
%
\section{The European VLBI Network}           
\label{sect:intro}

The European VLBI Network (EVN) consists of telescopes in Europe,
South Africa, and China. The network includes some of the largest 
telescopes in the world, for example the 100$\,$m Effelsberg antenna, 
the 76$\,$m Lovell Telescope, and the Westerbork Synthesis Radio Telescope.
There are in general three observing sessions
in a year. The data are normally correlated at
the EVN MkIV Data Processor at JIVE. Recent developments in the EVN 
({\it http:/$\!$/www.evlbi.org}) improved the data quality as well as the
operational reliability of the array, as summarised below.

\subsection{The Mark5 recording system}

As from 2004, all EVN stations are equipped with the Mark5 disk-based
recording system ({\it http:/$\!$/web.haystack.mit.edu/mark5/Mark5.htm}). 
This greatly improves the EVN reliability. Quick
fringe tests are possible by ftp-ing some data from the stations to the
correlator. The processing is done in an automated fashion using a
software correlator developed in the Kashima Space Research Center,
NICT (formerly CRL), Japan (Kondo et al. \cite{kon03}). With the Mark5 recording 
system it is also possible to send the data in realtime to the EVN MkIV   
Correlator via optical fibres. The first realtime EVN eVLBI map was made
on 26 April 2004 using a three-station network of telescopes including
Onsala, the WSRT and Jodrell Bank ({\it http:/$\!$/www.evlbi.org/evlbi/te017/te017.html}).
Recently a fourth antenna came on-line -- the 32$\,$m Torun telescope in Poland.

\begin{figure}
   \centering
   \includegraphics[angle=-90,width=120mm]{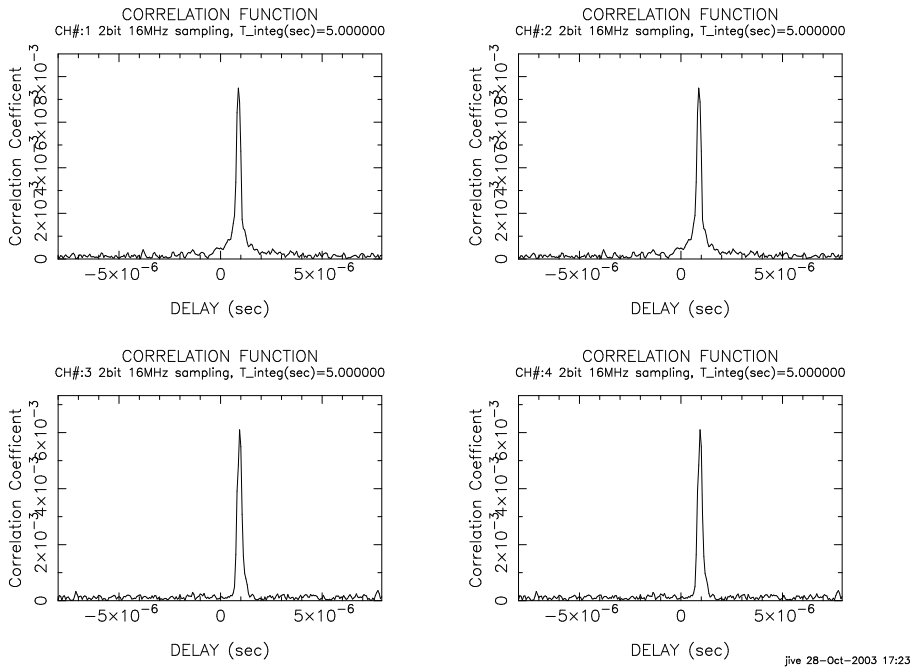}
   \caption{Fringes produced from data files ftp-d over the Internet.
            Ftp fringe-tests are very useful for early realisation of telescope
            failures. The data are processed by the NICT Software Correlator on
            a Cluster computer at ASTRON/JIVE. }
   \label{Fig:1}
\end{figure}

These technical developments provide not only a better calibrated and
more reliable array, but also the possibility of increasing the sensitivity.
From 2002, two headstack recording has been available allowing
512 Mbit/s data rate, but mechanical and logistical limitations meant it was used
rarely. With the Mark5 system, 1024 Mbit/s became a reality
in the EVN in 2004.

\subsection{Data calibration pipeline}

All experiments are processed by a data calibration pipeline (Reynolds,
Garrett \& Paragi \cite{rey02}). This 
is especially useful for users that have little experience with 
VLBI data reduction. This reduces the time needed for post-processing
significantly. The pipeline also produces rough maps of the calibrators,
and, upon request, the target source. It cannot be stressed how important
this may be for quick response science projects, to see whether the 
target is detected or not, without the user touching the data.

The pipeline results, the correlation FITS file, and
any information related to the observations are stored in the EVN Data Archive
({\it http:/$\!$/archive.jive.nl/scripts/listarch.php}). 

\subsection{Amplitude calibration}

The EVN is an inhomogeneous array with a variety of dishes with
different sizes and receivers with different characteristics.
In the past, $T_{\rm sys}$ measurements were sparse, and the calibration
diode temperatures were occasionally ill determined which caused difficulties 
in processing the data. 

Since 2002, the EVN introduced continuous $T_{\rm sys}$ monitoring.
The data are provided to the user in a format (so-called ANTAB file) 
which can be directly applied by the user without further editing.
In each observing session the calibration diode temperatures are
determined at the stations, and frequency dependent variations are
also corrected for.

\subsection{Correlator upgrade}

The field of view of VLBI images is usually limited to the order of 
100 milliarcseconds. This is because the data are averaged in frequency
and time during processing, to reduce the data size, in order to match
the capacity of the correlator and the post-processing hardware.
Currently the EVN MarkIV Correlator is being upgraded to be able to handle
sub-second integration times (1/4 s already possible) and an increased 
number of frequency channels. 

It is now possible to image a field of view of many arcminutes at 1.6~GHz 
(18$\,$cm), and the limitation is becoming the primary beam
(field of view) of phased arrays (like Westerbork) and big dishes
(Effelsberg, Lovell Telescope). Surveying projects of, for example nearby
galaxies ($< 10'$ in size) can be carried out in a single observation 
without observing with different pointings or recorrelating with different
positions.


\section{Faint source detection with the EVN}
\label{sect:Obs}

The thermal noise on VLBI images depends on the antenna sensitivities, the
noise characteristics of the receivers of each interferometer element, and
on additional noise sources (like the electronics and the target source itself).
The noise can be reduced by observing the target source longer or increasing
the recorded data rate. With tape recording, 256 Mbit/s was available using one
recording head, and this was extended to 512 Mbit/s using two recording heads
(see a two-head recording result in Fig.~2.).

With the Mark5 disk recording system, introduced at the EVN in 2004, it is
possible to record at 1024 Mbit/s. For two bit sampling and dual (RCP+LCP)
polarization this results in 128 MHz bandwidth in each polarization. The 
theoretical noise in the L- and C-bands (18$\,$cm and 6$\,$cm) at this high recording 
rate is about 10 microJy/beam with 5 hours on-source integration time.

Detecting sub-mJy sources requires a special observing mode, so-called 
phase-referencing. In this technique the target and a nearby calibrator 
($\sim$ 1~Jy source within 3--4 degrees) are  
observed alternately in short cycles. The atmospheric phase errors are corrected
for using the calibrator data. Phase-referencing is able to correct the phases
to some extent, but residual errors remain in the data. For this reason, a
phase calibrator is needed (order of 5--100 mJy, within $\sim$1 arcminute from the target) 
to reach the required noise level. Alternatively, one could use the summed
response of all detected sources in the field, so-called "full-beam" 
VLBI self-calibration (Garrett et al. \cite{MAG03}).
By correlating the observations with wide
field of view mode (see above), we have a better chance of having an in-beam
phase reference. To conclude, the EVN has the potential of detecting 40-50
microJy sources if the data are sufficiently well calibrated.

\begin{figure}
   \centering
   \includegraphics[width=100mm]{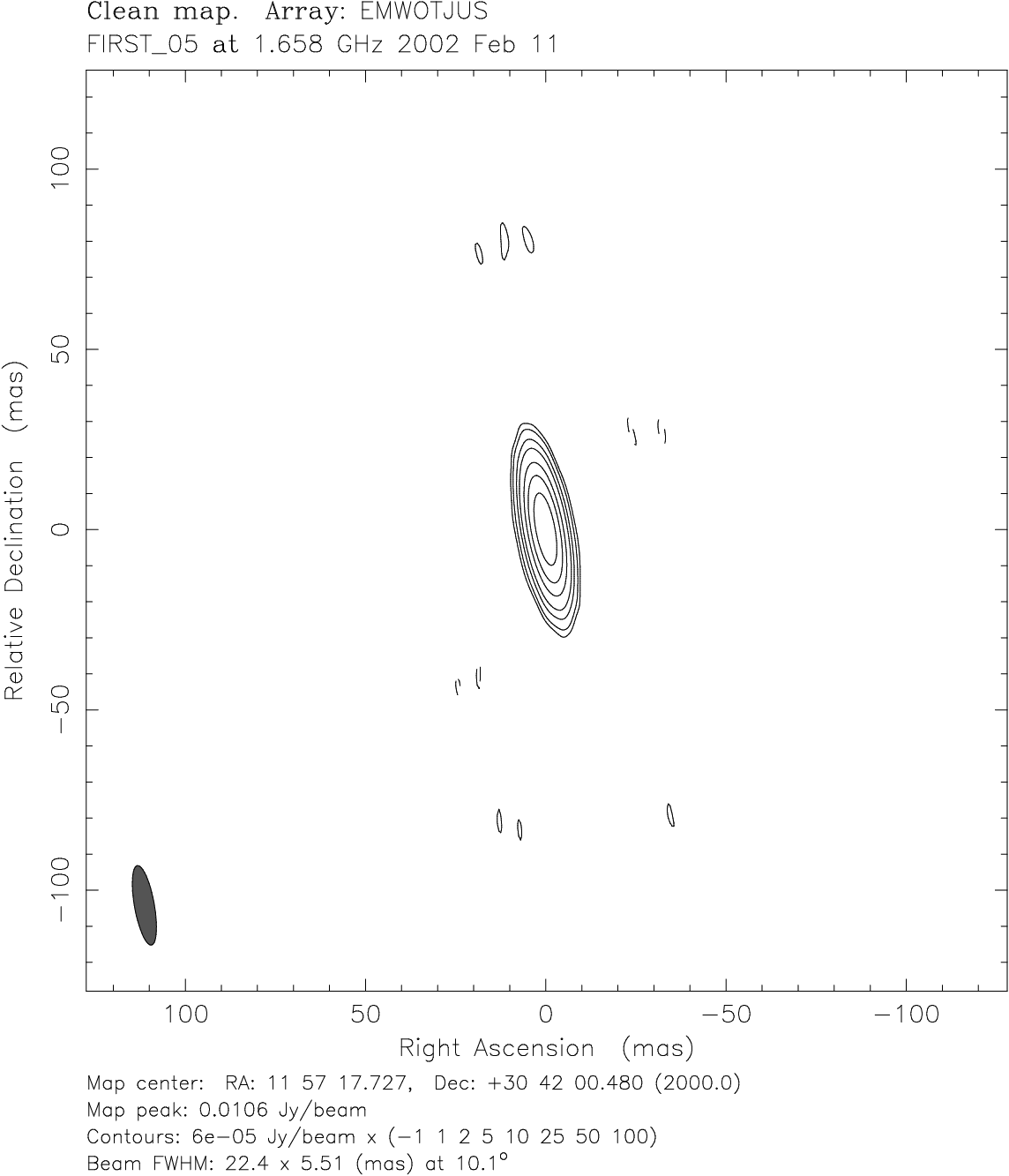}
   \caption{The 10 mJy source FIRST 051 observed in a phase-referenced
            Network Monitoring Experiment (Paragi et al. 2002). The rms noise on
            the map is 18 microJy/beam, achieved by 150 minutes on-source
            integration time with the EVN, using 512 Mbit/s recording rate.     
            }
   \label{Fig:2}
\end{figure}

\section{Microquasars and ULXs: potential EVN targets}

Some of the microquasars are permanently bright in the radio (like SS433), others
produce strong outbursts but are quite dim during quiescent periods (e.g. GRS 1915+105).
The majority of XRBs are not detected in the radio regime. Another consideration for Galactic sources
is that these are often variable on timescales of hours. With the EVN at 1 Gbit/s recording
rate, a 2 mJy target can be imaged at 100:1 dynamic range within only an hour of
observation. 

Potential candidates are the Ultra-Luminous X-ray (ULX) sources that may have radio counterparts.  
Some of these might be powered by Intermediate-Mass Black Holes (IMBH, Colbert \& 
Miller \cite{col-mill04}). Observation at milliarcsecond resolution could tell us whether these radio
sources have a compact jet structure or not. The existence of a jet would confirm the
presence of an accreting compact object in ULX sources. 

\begin{figure}
   \centering
   \includegraphics[angle=-90,width=120mm]{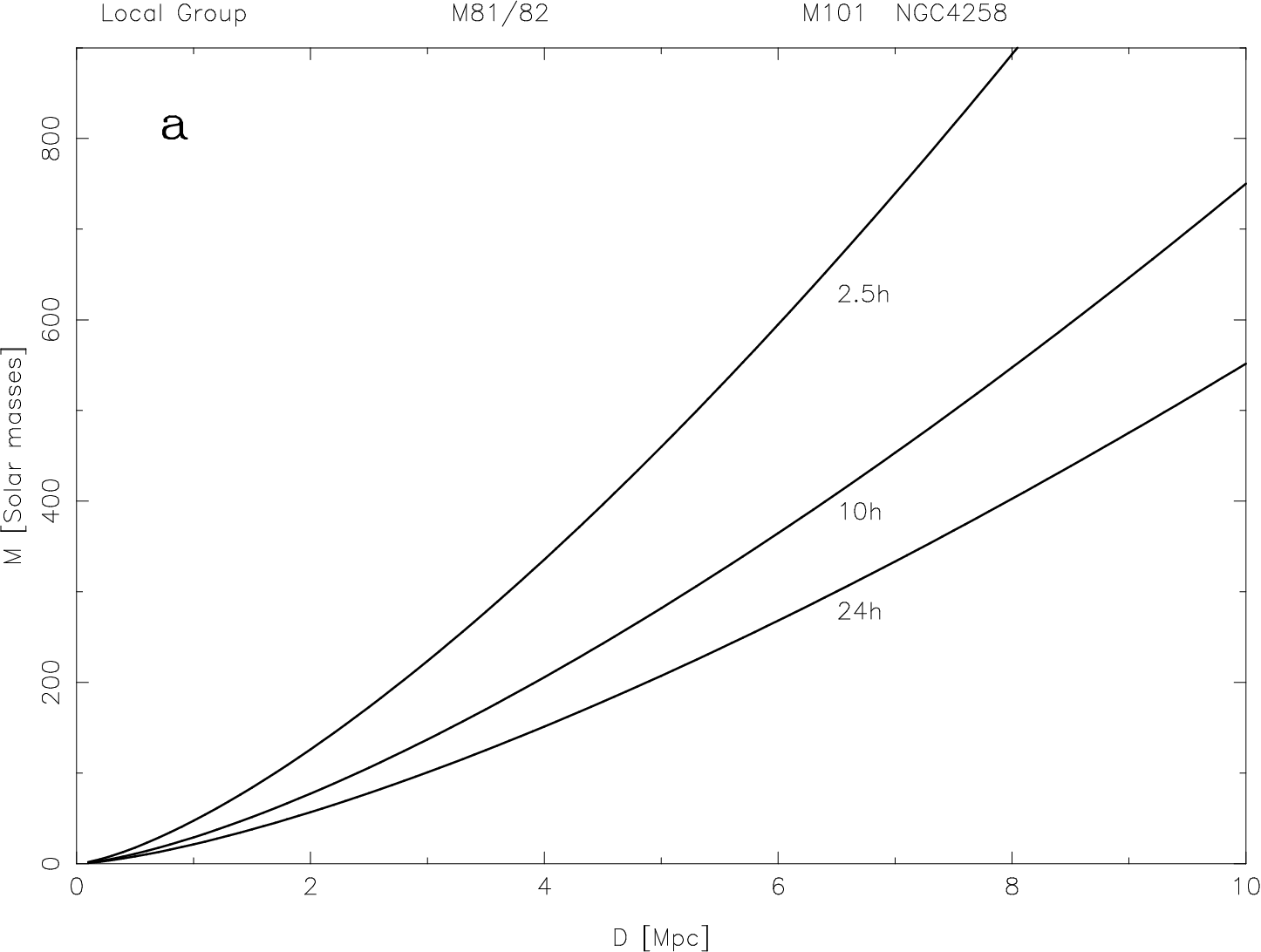}
   \includegraphics[angle=-90,width=120mm]{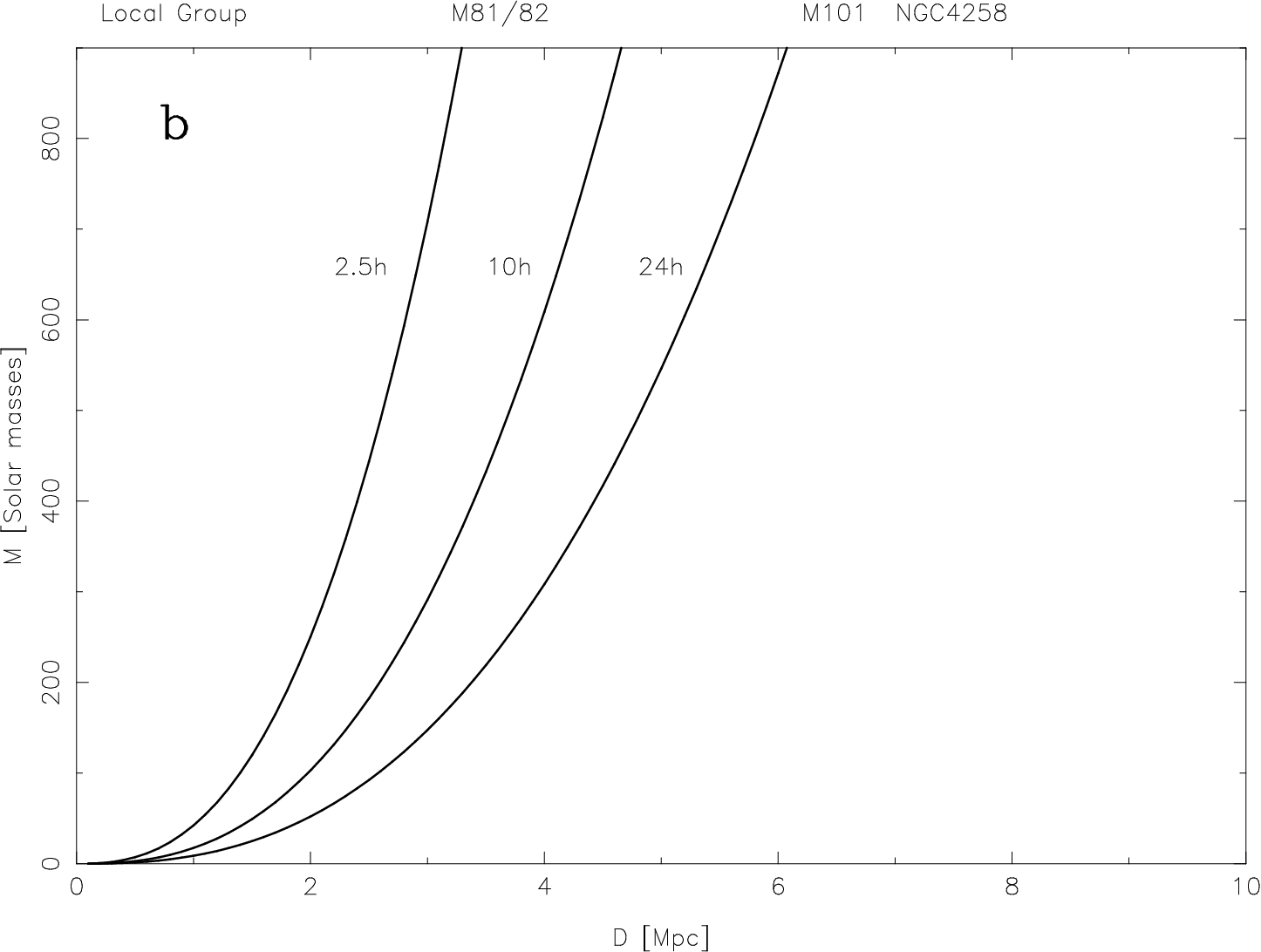}
   \caption{The minimum detectable mass of an IMBH (5 $\sigma$ detection at 5 GHz)
            in nearby galaxies, using different on-source integration times with
            the EVN. a) scaling up SS433 (300 mJy at 5 GHz, located at a distance of 5 kpc,
	    BH mass is about 10 M$_{\odot}$)
	    using $L_{\rm r}\propto M_{\rm BH}^{17/12}$ (Heinz \& Sunyaev \cite{hei-suny03}), and
             b) using the formula given by Maccarone (\cite{mac04}) and assuming an ULX luminosity of
	     $L_{\rm x}=10^{40}$ erg/s .
            }
   \label{Fig:3}
\end{figure}

One can speculate whether radio-loud IMBHs could be detected in nearby galaxies, if these
have a jet analogous to microquasars and active galactic nuclei. By scaling the radio flux 
density of SS433 (probably
hiding a 10 M$_{\odot}$ black hole), and assuming the $L_{\rm r}\propto M_{\rm BH}^{17/12}$ 
relation (Heinz \& Sunyaev \cite{hei-suny03}),
one could make such an estimate. On Fig.~3a
we plot the minimum detectable black hole mass with the EVN with different on-source integration
times. The results are shown with respect to the source distance, and some galaxies are
also indicated.

Note that there is no well established relation between the black hole mass and the radio 
luminosity. Heinz and Sunyaev (\cite{hei-suny03}) proposed $L_{\rm r} \propto M_{\rm BH}^{17/12}$ for 
optically thick flat spectrum core-dominated jets. Earlier Falcke and Biermann (\cite{fal-bier95}, 
\cite{fal-bier96})
came to similar conclusions, and found that this seems to be consistent with the observations.
Merloni et al. (\cite{mer03}) looked for correlation between $L_{\rm r}$, $L_{\rm x}$, and $M_{\rm BH}$,
and showed that the Heinz and Sunyaev (\cite{hei-suny03}) model is consistent with the observed properties of
Galactic and supermassive black holes if the accretion flows are radiatively inefficient.
Maccarone et al. (\cite{mac04}) achieved similar results but by excluding sources in the high/soft state.
Finally, Maccarone (\cite{mac04}) used the Merloni et al. (\cite{mer03}) correlation, 
$L_{\rm r} \propto L_{\rm x}^{0.6} M_{\rm BH}^{0.78}$, to express the radio flux of an  
accreting black hole system.
He showed that deep radio surveys may be useful for studying ULX sources.
In Fig.~3b we plot the minimum detectable black hole mass, corresponding to this latter 
relation. Both approaches agree that IMBHs could be detected in the closest galaxies with the EVN. 

\section{Conclusions}

The operational reliability and general performance of the EVN has improved greatly in the
past couple of years. New users not familiar with VLBI data reduction are particularly 
encouraged to use the EVN. A data calibration pipeline is available for users (especially 
useful for beginners), and support scientists are also available to support EVN users at
all stages of their observations. The sensitivity of the array has improved by a factor of two
recently with the introduction of the Mark5 recording system. Faint, variable Galactic sources
as well as extragalactic objects can be detected at the 40-50 microJy level.
According to our estimates, IMBHs, if they exist, can be safely detected with the EVN in nearby 
galaxies. 

\begin{acknowledgements}

We are grateful for useful suggestions from the anonymous referee.
The European VLBI Network is a joint facility of European, Chinese, 
South African and other radio astronomy institutes funded by their 
national research councils. Z.P. acknowledges the support of the 
European Community -- Access to Research Infrastructure and Infrastructure
Cooperation Networks (RadioNET FP5, contract No. HPRI-CT-1999-40003)
action of the Improving Human Potential Programme.

\end{acknowledgements}

\label{lastpage}


\begin{thebibliography}{99}

  \bibitem[2004]{col-mill04} Colbert E.J.M., Miller M.C., 2004,
     10th Marcel Grossmann Meeting on General Relativity,
     Rio de Janeiro, July 20--26, 2003. Invited talk. Eds. M. Novello,
     S. Perez-Bergliaffa \& R Ruffini. World Scientific, Singapore, 2005.
     (astro-ph/0402677)

  \bibitem[1995]{fal-bier95} Falcke H., Biermann P.L., 1995, 
     A\&A, 293, 665

  \bibitem[1996]{fal-bier96} Falcke H., Biermann P.L., 1996,
     A\&A, 308, 321

  \bibitem[2003]{MAG03} Garrett M.A., Wrobel J.M. \& R. Morganti, 2003,
     New Astron. Rev., 47, 385

  \bibitem[2003]{hei-suny03} Heinz S., Sunyaev R.A., 2003,
     \mnras, 343, 59

  \bibitem[2003]{kon03} Kondo T., Koyama Y. \& Osaki H., 2003,
     IVS CRL-TDC News No.23 (November 2003)

  \bibitem[2004] {mac04} Maccarone, T.J., 2004, 
     MNRAS, 351, 1049

  \bibitem[2003] {macetal03} Maccarone, T.J., Gallo, E. \& Fender, R.P., 2003,
     MNRAS, 345, L19

  \bibitem[2003]{mer03} Merloni A., Heinz S., di Matteo T., 2003
     \mnras, 345, 1057

  \bibitem[2002]{rey02} Reynolds C., Garrett M., Paragi Z., 2002,
     Proc. XXVII General Assembly of the International Union of Radio Science, 
     paper 924, session J8.P.4, URSI: Gent.

\end{thebibliography}
\end{document}